\documentstyle[11pt,paspconf,psfig]{article}

\begin{document}

\title{The Redshift Distribution of Distant Sources from Gravitational
Depletion in Clusters}

\author{Y. Mellier}
\affil{Institut d'Astrophysique de Paris and Observatoire de Paris
DEMIRM, France.}




\begin{abstract}
Gravitational lensing can  be used to analyze the
redshift distribution of faint galaxies.  In particular the 
magnification bias modifies locally the galaxy number density of 
lensed sources observed in lensing clusters. This {\sl depletion 
area} probes 
 the redshift distribution of galaxies beyond $B=25$. In this
proceedings I present this new tool to infer the redhsift distribution
of faint galaxies.
\end{abstract}




\section{Introduction}

With the coming of 10 meter
class telescopes equipped with wide field multi-object
spectrographs, deep redshift  surveys will be extended to thousands of galaxies
 and will permit 
to explore in detail the evolution of clustering of galaxies,
the history of star formation up to $z=4$ for galaxies with $B \le 25$.\\
The study of galaxies with magnitudes $B>25$ are also important 
for the models of galaxy formation :   
 we do not know yet whether they are all at large
redshift or if there is a significant fraction of faint nearby dwarfs
galaxies.  The knowledge of their redshift distribution is 
 also necessary for mass reconstruction 
using lensing inversion, and can  
  be a major source of uncertainty in the mass
determination for the most distant lensing clusters 
(see Luppino \& Kaiser 1997). Bernardeau et al (1997) and Jain \&
Seljak (1997) have  
 emphasized that even the
 study the large-scale mass distribution using
 weak lensing  need the redshift distribution of
the faintest galaxies, because the variance and the skewness of the
magnification strongly depends of the redshift of the lensed background
sources.

Unfortunately, 
beyond  $B=25$, even 10 meter class telescopes are unable to provide   
redshifts of a complete sample of galaxies.
The possibility of using photometric
redshifts which was proposed by the beginning of eighties
 is now re-investigated in great details. But observations
as well as reliability tests are still underway (Connolly et al. 1995. 
Since they are based on
theoretical evolution scenarios of galaxies, their predictions about
faintest galaxies are not fully confirmed yet. Furthermore, there is no
hope to calibrate the photometric redshifts of the faint samples
 with spectroscopic data.

An attractive alternative to spectroscopy consists in using of the
magnification
and distortion effects induced by gravitational lensing on extended
objects. In particular, the magnification bias can eventually produces 
 depletion areas in the projected galaxy number density of 
background sources observed in rich clusters 
 whose size and shape depend on their redshift distribution. In the 
following section I present the basic principle of the technique and
first results.

\section{The distribution of faint galaxies from the
magnification bias}
The projected number density of galaxies through a lensing cluster 
at radial distance $r$ from the cluster center and with
magnitude lower than $m$, $ N(<m,r)$, results from
 the competition between the gravitational magnification that increases
the detection of individual objects and the deviation
of light beam that increases the area and thus decreases the apparent
number
density.  Therefore the amplitude of the magnification bias
depends on the slope of the galaxy counts, $\gamma$, as a function of magnitude and
on the magnification factor of the lens (Broadhurst et al. 1995):
\begin{equation}
   N(<m,r) = N_0(<m) \ \mu(r)^{2.5\gamma-1} \ ,
\end{equation}
 where  $\mu(r)$ is the magnification factor of the lens, $N_0(<m)$ the
intrinsic number density in a nearby empty field and
  $\gamma$  is the intrinsic count slope:
\begin{equation}
\gamma = {dlogN(<m) \over dm }\ .
\end{equation}

when the slope is higher than $0.4$ the number density increases,
whereas below $0.4$ is decreases and the radial distribution shows a
typical depletion curve (see Figure 1).

When the slope is lower than $0.3$, a sharp decrease of the number of
galaxies
is expected
 close to the critical radius of the lens corresponding to the redshift
of the background sources. For a broad redshift distribution,
 it can result a shallower depletion between the smallest and
the largest critical line
which depends on the redshift  distribution of the galaxies
(Figure 1).  Therefore, the analysis of the shape of the depletion
curves provide a new way
to sort out  their redshift
distribution. As the lensing inversion, this is a statistical
 method which can also
 infer redshift of very faint sources (up to $B=28$) but does not need
anymore information on the shapes of arclets. However, the need of a
good
lens modeling is still necessary.

 This method was first used by Fort et al (1997) in the cluster
Cl0024+1654 to study the
faint distant galaxies population in the extreme range of magnitude
$B=26.5-28$ and $I=25-26.5$. For these selected bins of magnitude they
found on their CFHT blank fields
that the counts slope was near 0.2, well suited for the study of
the  effect. After analysis of the shape of the depletion
curve (figure 4), $60\% \pm 10\%$ of the
$B$-selected galaxies were found
between $z=0.9$ and $z=1.1$ while most of the remaining $40\% \pm 10\%$
galaxies appears to be broadly distributed around a redshift of
$z=3$. The
$I$ selected population present a similar distribution  with two maxima,
but
spread up to a larger redshift range  with about 20\% above  $z > 4$
(Figure 1).

This first tentative must be pursued on many lensing clusters in
order to
provide significant results on the redshift distribution of the faintest
distant galaxies. Though it is a very promising approach, it also need
to be
applied on clusters with simple geometry. Furthermore, the detection
procedure demands ultra-deep exposures with subarcsecond seeing.

\begin{figure}
\psfig{figure=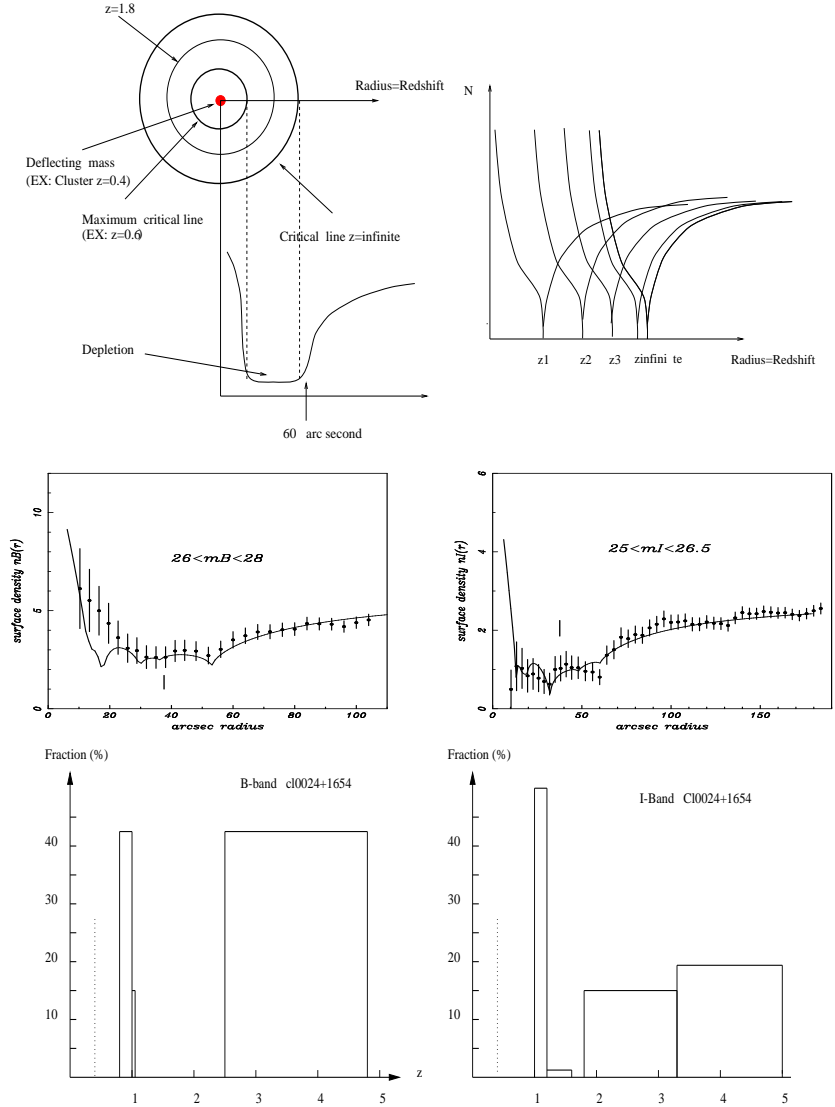,height=14.7 cm}
\caption{Depletion by a singular isothermal sphere
as it would be observed on the sky
 and radial density of galaxies (top left).
For a given redshift,
the minimum of the depletion is sharp and its radial position is
equivalent to a redshift (top right). The minimum increases with the
redshift of sources but the depletion curves tighten and converge
towards the curve corresponding to sources at infinity.  In a realistic
case, the redshift distribution is broad and the individual curves must
be added. In this case, instead of
the single peaked depletion we expect a more pronounced minimum  between
two radii (i.e. two redshifts; top left). The middle panels show the
depletion curves observed in $B$ and $I$ in Cl0024.
 Since  the mass distribution of this lens is well known, one can
recover the
redhsift of the sources for the $B$ and $I$ populations (bottom panels:
note that this is a fraction of galaxies. The width of boxes is
the redshift range, not a total number of galaxies).
}
\end{figure}

%
%

%
%

\section{Conclusions }
The redshift distribution of galaxies beyond $B=25$ is a crucial
scientific question for galaxy evolution and weak lensing studies for
mass reconstruction.
 The depletion curves of galaxy number density produced by  
magnification bias is an  innovative way which can probe 
the redshift distribution of galaxies as faint
as $B=28$. The first tentative by Fort et al (1997) 
 demonstrates that depletion curves can be observed in Cl0024+1654 and
A370. However, a good modeling of the lensing clusters is needed in
order to infer the redshift distribution of the lensed sources.  This 
method is still at its infancy and the first results 
 are questionable. Hence, it 
 must be considered jointly with other techniques like 
photometric redshifts or lensing inversion (Kneib et al 1994, 1996).  \\
Whatever
the method, how can we be sure that these redshifts obtained
from non-standard and indirect techniques are
correct ?
 Preliminary deep spectroscopic and 
 multicolor photometric surveys of
arclets show that the faintest galaxies seem to have a redshift
distribution like the ones predicted by Fort et al (Pell\'o, private
communication). But this
key issue demands ultra-deep CCD spectroscopic exposures with
the VLTs.  This should be in the future a major challenge  
for the gravitational telescopes.

\section*{Acknowledgments}
I thank F. Bernardeau, B. Fort, R. Ellis, R. Pell\'o, P. Schneider and 
L. Van Waerbeke for
  stimulating discussions.
 This work was supported by the Programme National de Cosmologie.
\begin{figure}
\centerline{\psfig{figure=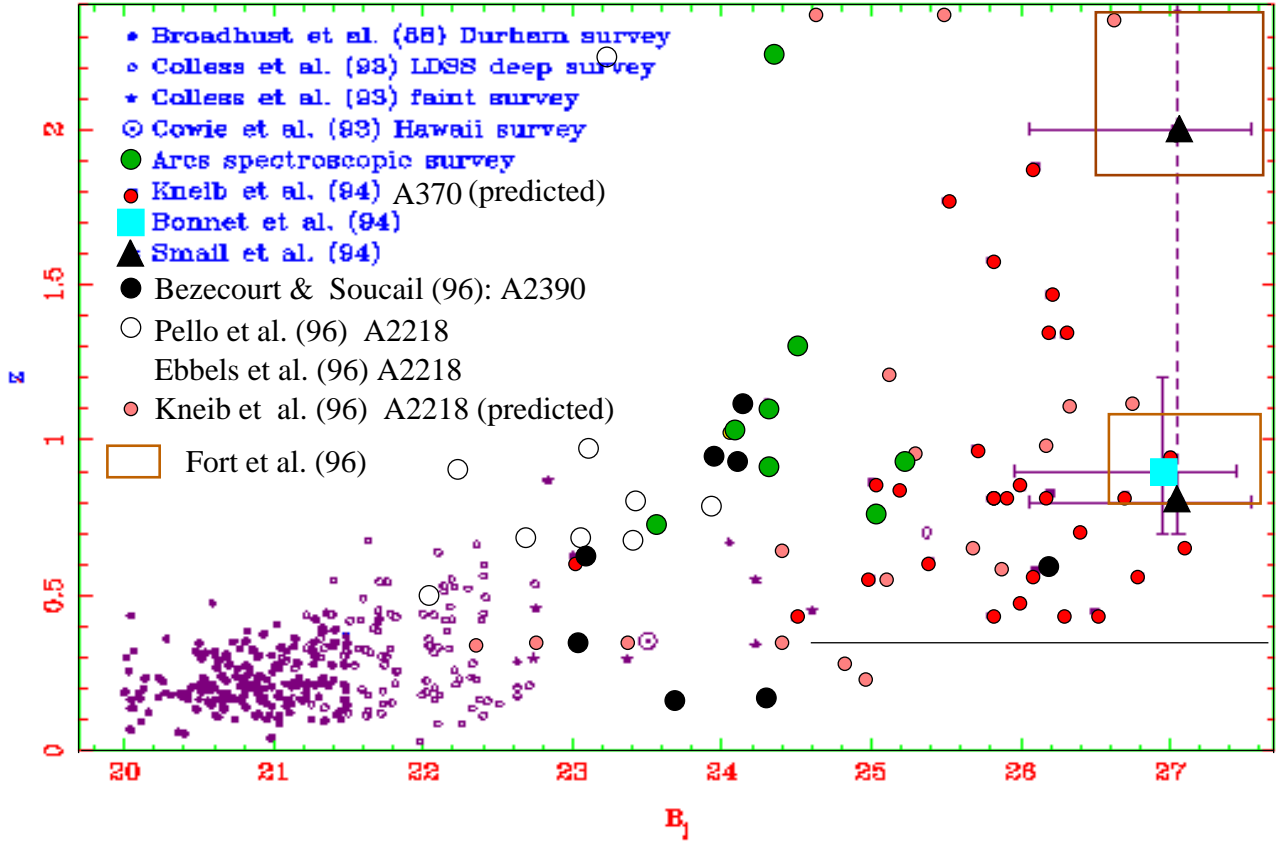,height=12.0 cm}}
\caption{A magnitude-redshift diagramme showing the positions of
the redshift surveys (dark symbols on the left), the arc(let)s
spectroscopic surveys (large circles. Those concerning A2218
have been kindly provided by Pell\'o prior to publication), the
predictions
of lensing inversions for A370 and A2218 (small circles),
of weak lensing studies  by Bonnet et al. and Smail et al. (triangles)
and finally, of the depletion curves in Cl0024 (large boxes). The
spectroscopic redshift of Cowie et al. (1996) with Keck would be between
$B=22.5$ and $B=24.5$. We see the potential interest of gravitational
lensing which provide redshifts up to $B=28$.  The straight line on
bottom right is the redshift of A370 which is a limit of the lensing
inversion in this cluster.
}
\end{figure}

%


\begin{references}
\reference F. Bernardeau, L.  van Waerbeke, Y.  Mellier, 1997, A\&A 322,
1.
\reference  T. J. Broadhurst, A. N. Taylor, J. Peacock, 1995, ApJ 438,
49.
\reference A. Connolly, I. Csabai, A. S. Szalay, D. C. Koo, R. G. Kron,
J. A. Munn 1995, AJ 110, 2655.
\reference B. Fort, Y. Mellier, M. Dantel-Fort 1997, A\&A 321, 353.
\reference B. Jain, U. Seljak 1997, ApJ 484, 560.
\reference J.-P. Kneib, G. Mathez, B. Fort, Y. Mellier, G. Soucail,
P.-Y. Longaretti 1994, A\&A 286, 701.
\reference J.-P. Kneib, R. S. Ellis, I. Smail, W. J. Couch, R. M.
Sharples 1996, ApJ 471, 643.
\reference G. Luppino, N. Kaiser 1997, ApJ 475, 20.
\end{references}
\end{document}